\documentclass[sigconf]{acmart}
\usepackage{tablefootnote}
\AtBeginDocument{%
  }

\setcopyright{acmlicensed}
\copyrightyear{2026}
\acmYear{2026}
\acmDOI{}
\acmConference[CSCW '26 Workshop]{Women's Health as Cooperative Work}{2026}{Salt Lake City, UT, USA}
\acmISBN{}

\begin{document}

\title{Women's Genetic Health as Family Cooperative Work: Empirical Findings and a Case for Participatory AI}

\author{Ziqi Yang}
\email{ziqiy30@uci.edu}
\affiliation{%
  \institution{University of California, Irvine}
  \department{Department of Informatics}
  \city{Irvine}
  \state{California}
  \country{USA}
}

\author{Yunan Chen}
\email{yunanc@ics.uci.edu}
\affiliation{%
  \institution{University of California, Irvine}
  \department{Department of Informatics}
  \city{Irvine}
  \state{California}
  \country{USA}
}

\author{Daniel A. Epstein}
\email{epstein@ics.uci.edu}
\affiliation{%
  \institution{University of California, Irvine}
  \department{Department of Informatics}
  \city{Irvine}
  \state{California}
  \country{USA}
}

\renewcommand{\shortauthors}{Yang et al.}

\begin{abstract}
Genetics play a significant role in women's health, with family members often experiencing shared symptoms or hereditary risks across generations.
Yet health technologies, including emerging health AI tools, have supported women's health as an individual health practice.
Drawing on a survey of 249 U.S.\ adult women, we characterize women's genetic health information work as cooperative work distributed across intergenerational family networks, identify who does this work and why, and document the barriers that fragment it.
We find that generative AI use is a strong predictor of genetic health perception, but current tools focus on individuals rather than supporting collective coordination.
We argue that designing AI for this domain requires participatory approaches that engage diverse women, healthcare professionals, and legal and community stakeholders, and we bring open design questions to the workshop.
\end{abstract}


\settopmatter{printacmref=false}
\maketitle

\noindent\textbf{Workshop:}
CSCW 2026 -- Women's Health as Cooperative Work:
Participatory AI Across Care, Community, and Policy
(October 11, 2026).

\par\smallskip
\noindent\textbf{Proceedings index:}
\url{https://arxiv.org/abs/2611.03776}

\par\smallskip
\noindent\textbf{Assigned report number:}
\texttt{CSCW-WHCW-2026-04}

\par\medskip

\section{Introduction}

Genetics contribute significantly to people's health through inheritance across generations, covering aspects from everyday wellbeing to hereditary disease~\cite{yang2026sharing}.
Women's health experiences are particularly shaped by genetic factors: genetic resemblance results in similar timing of menarche and menopause and shared menstrual symptoms between sisters~\cite{dvornyk2012genetics,laven2015genetics}, while genes such as BRCA1 and BRCA2 elevate risks of breast and ovarian cancer that run in families~\cite{evans2020population,witt2010genetics}. Because biological family members across multiple generations often share similar health experiences, and family communication has also been a crucial source of information for genetic health and health literacy~\cite{corona2012talking,zambrana2015association}, information from families could greatly support women in understanding and managing their own health.

HCI has focused on technologies for women to manage their health information, but these design approaches have largely overlooked health experiences impacted by these familial genetics~\cite{yang2026sharing}.
Self-tracking tools have supported women in documenting personal health around menstruation, pregnancy, and menopause~\cite{epstein2017examining,figueiredo2017selftracking,lopezburst2024menopause}; AI has been used to help women interpret personal health data or seek information~\cite{Zhou2026It, Kaleva2026Privacy}. Yet, these technologies are typically designed for individual use.
Where family health technology exists, it tends to center on the needs of one member with an acute or chronic condition, while other members act as supporters~\cite{pina2017personal,lu2024unpacking}.
There is therefore a gap in understanding how technology can support families in building, maintaining, and passing down their shared genetic health history.

In our position paper, we summarize findings relevant to cooperative work and participatory AI from our survey of 249 U.S.\ adult women around genetic health~\cite{yang2025family, yang2026sharing}. We found that participants shared various women's health experiences mostly through opportunistic in-person conversations. However, the families shared limited details due to not only a lack of awareness of genetic health, but also gaps in time and space, family tensions, and cultural, generational, and gendered stigma.
Generative AI use is among the strongest predictors of engagement, suggesting an potential for these tools alongside a risk of mismatch, as AI designed for individual retrieval cannot address the cooperative, multi-stakeholder nature of this work~\cite{yang2025family}.
We draw on this survey to characterize women's genetic health sharing as cooperative work distributed across intergenerational family networks, and bring open questions on designing participatory AI for women's genetic health to the workshop.

\section{A Survey of 249 U.S. Women}

To examine families' practices and challenges around sharing women's genetic health experiences, we conducted an online survey with 249 adult women in the United States (ages 18-84) in summer 2025, recruited through Prolific using stratified sampling to approximate the U.S. age distribution~\cite{yang2026sharing,yang2025family}.
The survey covered participants' genetic health knowledge and perceptions, their practices of learning and sharing women's health experiences within families, and their use of health technologies, and was analyzed through reflexive thematic analysis and ordinal logistic regression.


\subsection{What, When and Why Families Share}

We found that families shared women's genetic health information across domains from everyday life to chronic conditions and diseases, including menstrual patterns and symptoms, pregnancy and fertility experiences, menopause, and chronic conditions or illnesses.
Menstrual patterns were the most widely exchanged: 75.9\% of participants reported learning about their family's menstrual histories and 75.5\% had shared their own, typically when a daughter reached menarche and mothers or grandmothers disclosed that heavy periods or irregular cycles ``\textit{run in the family.}''
Pregnancy and fertility histories were discussed by 62.7\%.
Hereditary conditions were known to 55\%, though knowledge was often fragmentary: ``\textit{My aunt says my mother had breast cancer, but she doesn't know anything more}'' (P88, 45 yrs).

Sharing typically occurred through in-person conversations during family gatherings and life transitions.
The mother-daughter dyad was most frequently mentioned. 90.8\% of participants reported learning from their mother, while only 8.4\% from their father.
Grandmothers, aunts, and siblings also shared experiences relevant to genetics.
Participants also mentioned benefits and motivations for sharing, including \emph{preparing} younger family members for health transitions they had not yet experienced; \emph{validating} one's own experiences by locating them within a family pattern; and collectively \emph{making sense} of recurrent conditions.


\subsection{The Role of Health Technologies}

\begin{figure}[h!]
    \centering         
\centering
        \includegraphics[width=0.8\linewidth]{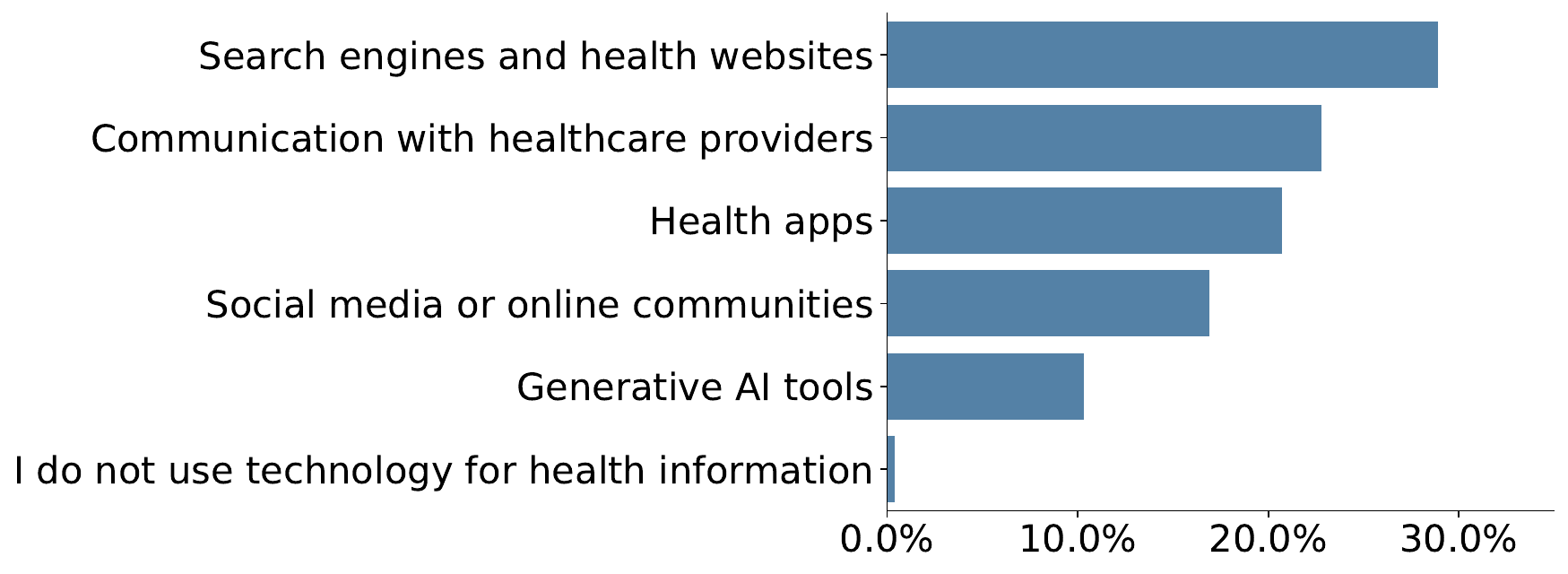}
        \caption{Types of technologies that participants used to learn and manage health information. Search engines and health websites were the most commonly used resources, followed by communication with healthcare providers and health apps.}
        \Description{This figure shows a horizontal bar chart of technologies that participants used to learn and manage health information. The bar for “Search engines and health websites” is the longest at a little over 30 percent. “Communication with healthcare providers” follows at roughly 25 percent. “Health apps” are used by about 23 percent of participants. “Social media or online communities” is close to 18 percent. “Generative AI tools” are used by around 10 percent. The shortest bar, just above zero, represents those who reported “I do not use technology for health information.}
        \label{fig:health-management-tools}
\end{figure}

Participants used technologies such as health websites, apps, and medical records mainly for personal health rather than sharing with family (Appendix \ref{fig:health-management-tools}). Sharing beyond conversations was mentioned mostly as ad hoc practices (texted photos, blog links, or genetic testing results) and no participant described systematic family health documentation. Participants expressed interest in AI, such as surfacing patterns by "\textit{[showing] whether something is passed down}"(P40), "\textit{[enhancing] effective conversations between women and expert opinions}"(P25), or being an "\textit{AI interviewer}" for documentation(P83).
A few raised concerns towards AI, mentioning "\textit{medical info should not be used to train an AI"}(P206).
Generative AI use was also the strongest predictor of perceived genetic health importance and impact among individual categories of technologies. Women who used it were 2.45 times more likely to perceive family history as important and nearly twice the odds of perceiving strong genetic impact (Appendix \ref{tab:rq2_base}). Yet, they are mostly used for individuals' health Q\&A.



\subsection{Challenges in Sharing Across the Family}

Despite these motivations, participants described barriers that fragmented the sharing of their health experiences.
Many shared limited details due to a lack of awareness of genetic health, making it difficult to know what information was worth sharing or seeking.
Sharing was largely opportunistic, often triggered by a diagnosis, a pregnancy, or a health crisis.
Gaps in time and space also disrupted the sharing . This included conditions where family divorce, relationship tension, and the deaths of older relatives.
Lastly, we identified three forms of stigma that made sharing challenging.
\emph{Cultural stigma} treated reproductive and menstrual health as private and taboo, with ethnic and religious contexts intensifying this norm.
\emph{Generational stigma}, described as "generational gap" or "generation silence", where older relatives' trivializing symptoms or relectance to share resulted in a lack of sharing.
\emph{Gendered stigma} excluded male family members from these conversations. Only 8.8\% of participants had shared genetic health information with their father, leaving the paternal side of the genetic tree absent from the current sharing network.

Women's knowledge and perceptions towards genetic health also varied with their family circumstances and reproductive experiences.
Participants who reported contact with larger numbers of biological family members were 44\% more likely to rate family health history as important.
Having given birth and having a daughter were also associated with higher perceived genetic impact and knowledge.
These patterns suggest that women's engagement with genetic health is unevenly shaped by family circumstances, reproductive experiences, and life stage.



\section{Discussion and Workshop Participation}


Our findings surfaced that women's genetic health information work could also be seen as cooperative work, spread across the family rather than one person, where family members coordinate to make sense of genetic health together~\cite{schmidt1992taking,ackerman2000intellectual}, which shifts the focus from the individual woman to the family as a whole.
Yet our findings suggest that this work is not currently distributed equally among family members.
Gendered stigma influences who gets included and excluded, and distance, tensions in relationships, and the death of older relatives make information harder to pass down generations.
This leads to issues such as the limited engagement for paternal side of the family, and the burden of including remote relatives falling on whoever happens to coordinate this work.
We also find that existing health technologies has not been sufficient, as they are designed for individual use rather than for a family to build and maintain a shared record together.
We therefore suggest that future technology could better support this cooperative work by sharing the work of inclusion and coordination among family members, rather than concentrating it on individuals.

Generative AI use was among the strongest predictors of genetic health engagement in our survey, and participants imagined AI helping surface whether a condition runs in the family.
This points to opportunities for AI to support cooperative work, such as connecting fragments of knowledge scattered across family members, coordinating sharing across generations and time gaps, and offering indirect ways to raise topics a family member might otherwise avoid.
At the same time, our findings suggest the need to design \emph{with} families rather than only \emph{for} them.
An AI that only answers one person may miss knowledge spread across a family, fail to support coordination over time, or mishandle consent when its answers involve other members; built without input from affected families, it could also repeat the stigma patterns we found.
Families may further disagree on who owns or manages genetic health information, a tension better resolved by involving affected people directly in design than by adding more settings and options~\cite{delgado2023participatory}.


Based on our findings~\cite{yang2026sharing,yang2025family}, we bring some questions to the workshop for discussion. We are particularly interested in: \textit{How can we design AI to support women in families to collaboratively manage their genetic health, and how can we invite women to address challenges such as privacy risks or stigma?} Smaller questions follow from this. For example, how can we build a family health repository that can be used for generations without burdening a particular member? Whose consent should matter when AI-generated insights involve distant relatives, and who can decide on information access? And how might we leverage AI to engage members beyond whoever currently coordinates this work given the existing family tensions or sociocultural norms in women's health?
We hope to develop these questions with participants whose work spans care coordination, feminist design methods, participatory AI governance, and legal policies.

\bibliographystyle{ACM-Reference-Format}
\bibliography{references}

\newpage
\appendix
\onecolumn
\section{Appendix: Survey Results}

\subsection{Descriptive Survey Result for Health Technology Use}

\subsection{Association between GenAI Use and Perceived Genetic Health Knowledge and Perception} 
\begin{table}[htbp]
\centering
\caption{Survey Questions for Genetic Health Knowledge and Perceptions}
\label{tab:dependent_variables}
\begin{tabular}{|p{4cm}|p{7cm}|p{4cm}|}
\hline
\textbf{Variable} & \textbf{Survey Question (Text)} & \textbf{Response Options} \\
\hline

Genetic Health Knowledge &
How much do you think you know about how genetics affect women’s health? &
5-point Likert scale (1--5) \tablefootnote{
1 = Not at all knowledgeable;
2 = Slightly knowledgeable;
3 = Moderately knowledgeable;
4 = Knowledgeable;
5 = Very knowledgeable}\\
\hline

Perceived Importance of Family Health History &
How important do you think it is to know about your family’s health history, especially related to women’s health issues?&
5-point Likert scale (1--5) \tablefootnote{
1 = Not at all important; 
2 = Slightly important; 
3 = Moderately important; 
4 = Important;
5 = Very important.} \\
\hline

Perceived Genetic Impact &
How much do you think genetics affects women’s health conditions among all influencing factors?& 
5-point Likert scale (1--5) 
\tablefootnote{ 
1 = No impact; 
2 = Slight impact; 
3 = Moderate impact; 
4 = High impact; 
5 = A lot of impact}\\
\hline

\end{tabular}
\end{table}

\begin{table}[htbp]
\centering
\caption{Survey Question for Digital health technology engagement}
\label{tab:iv_rq2}
\begin{tabular}{|p{2.5cm}|p{6cm}|p{6.8cm}|}
\hline
\textbf{Variable} & \textbf{Survey Question (Text)} & \textbf{Response Options} \\
\hline

Health Information Technology Use&
What types of technology do you use to find or learn about health information? (Select all that apply) &
Search engines and health websites; Generative AI tools; Health apps; Social media or online communities; Communication with healthcare providers; I do not use technology for health information; Other\\
\hline


\end{tabular}
\end{table}

\begin{table}[htbp]
\centering
\caption{Technology Models: Associations with Self-Reported Genetic Health Knowledge, Perceived Importance of Family Health History, and Perceived Genetic Impact on Women’s Health}
\label{tab:rq2_base}
\begin{tabular}{|l|c|c|c|}
\hline
 & \textbf{Knowledge} & \textbf{Importance} & \textbf{Impact} \\\hline
\textbf{Predictor} & OR (95\% CI) & OR (95\% CI) & OR (95\% CI) \\
\hline

Technology Breadth& 1.25* (1.02--1.52) 
& 1.47*** (1.17--1.85) 
& 1.27* (1.04--1.56) \\\hline

Generative AI Use 
& 1.62 (0.90--2.91) 
& 2.45** (1.31--4.61) 
& 1.92* (1.13--3.28) \\

\hline
\multicolumn{4}{|l|}{\footnotesize Models adjusted for age, education, race, family size, and reproductive variables.} \\\hline
\multicolumn{4}{|l|}{\footnotesize * $p<.05$, ** $p<.01$, *** $p<.001$.}
\\ \hline\end{tabular}
\end{table}


Among individual technology categories, use of generative AI tools was significantly associated with stronger endorsement of both perceived importance of family health history (OR = 2.45, $p<.01$) and perceived genetic impact (OR = 1.92, $p<.05$). 

\end{document}